\documentclass[12pt,amsart]{iopart}
\usepackage{fullpage,graphicx,epsf}
\usepackage{iopams}%
\usepackage{amssymb}
\usepackage{amscd}
\usepackage{cite}
\usepackage{color}%
\usepackage{graphicx}
\usepackage{epsfig}
\usepackage[english]{babel}
\usepackage{amsfonts}
\begin{document}

\def\be{\begin{equation}}
\def\ee{\end{equation}}
\def\bea{\begin{eqnarray}}
\def\eea{\end{eqnarray}}
\def\bef{\begin{figure}[h!]}
\def\eef{\end{figure}}
\def\bml{\begin{mathletters}}
\def\eml{\end{mathletters}}
\def\l{\label}
\def\b{\bullet}
\def\no{\nonumber}
\def\fr{\frac}
\def\th{\theta}
\def\eps{\epsilon}
\def\o{\omega}
\def\O{\Omega}
\def\p{\partial}
\def\ph{\phi}

\title{Quasistationarity in a model of classical spins with long-range interactions}
\author{Shamik Gupta and David Mukamel}
\address{Physics of Complex Systems, Weizmann Institute of Science, Rehovot 76100, Israel}
\ead{shamik.gupta@weizmann.ac.il, david.mukamel@weizmann.ac.il}

\begin{abstract}
Systems with long-range interactions, while relaxing
towards equilibrium, sometimes get trapped in long-lived
non-Boltzmann quasistationary states (QSS) which have lifetimes that
grow algebraically with the system size. Such states have been
observed in models of globally coupled particles that move under
Hamiltonian dynamics either on a unit circle or on a unit spherical
surface. Here, we address the ubiquity of QSS in long-range systems
by considering a different dynamical setting. Thus, we consider an
anisotropic Heisenberg model consisting of classical Heisenberg
spins with mean-field interactions and evolving under classical spin
dynamics. Our analysis of the corresponding Vlasov equation for time 
evolution of the phase space distribution shows
that in a certain energy interval, relaxation of a class of 
initial states occurs over a
timescale which grows algebraically with the system size. We support
these findings by extensive numerical simulations. This work further
supports the generality of occurrence of QSS in long-range systems
evolving under Hamiltonian dynamics.
\end{abstract}
\date{\today}
\maketitle

\section{Introduction}
Long-range
interacting systems have attracted considerable interest in recent
years \cite{review1,review2,review3,review4,review5,review6}. These systems
are characterized by an
interparticle potential which in $d$ dimensions decays at large
separation, $r$, as $1/r^{\alpha}$, with $\alpha \le d$. Examples
include non-neutral plasmas \cite{Nicholson:1992}, dipolar
ferroelectrics and ferromagnets \cite{Landau:1960}, self-gravitating
systems \cite{Paddy:1990}, two-dimensional geophysical
vortices \cite{Chavanis:2002}, etc. Long-range interactions lead to
non-additivity, whereby thermodynamic quantities scale superlinearly
with the system size. This may result in equilibrium properties which
are unusual for short-range systems, e.g., a negative microcanonical
specific heat \cite{Lynden-Bell:1968,Thirring:1970}, inequivalence of
statistical ensembles \cite{Barre:2001,Mukamel:2005}, and many
others \cite{Bouchet:2005}.

As for the dynamical properties, long-range systems often exhibit broken
ergodicity \cite{Mukamel:2005,Bouchet:2008} and intriguingly slow
relaxation towards equilibrium \cite{Mukamel:2005, Chavanis:2002,Ruffo:1995,Yamaguchi:2004,Campa:2007,Joyce:2010}.
Such slow relaxation has been discussed in the context of self-gravitating systems (see, e.g., \cite{Levin:2010,Gabrielli:2010}), 
and has recently been demonstrated in a model of globally
coupled particles (inertial rotors) moving on a unit circle and evolving under
deterministic Hamilton dynamics. In this so-called Hamiltonian
mean-field (HMF) model, it has been found that for a wide
class of initial distributions, the relaxation time diverges with
the system size \cite{Ruffo:1995}. For some energy interval, the
relaxation proceeds
through intermediate long-lived quasistationary states (QSS). These
non-Boltzmann states exhibit slow relaxation of thermodynamic
observables, and have a lifetime which grows algebraically with the
system size \cite{Yamaguchi:2004,Jain:2007}. As a result, the system
in the thermodynamic limit never attains the Boltzmann-Gibbs equilibrium
but remains trapped in the QSS. At other energies, however,
relaxation occurs faster on a time which grows logarithmically with
the system size. Quasistationary states in the HMF model
exhibit interesting features
like anomalous diffusion and non-Gaussian velocity
distributions \cite{Latora:1999,Latora:2001,Bouchet:2005R}. Generalization of the model to include
anisotropy terms in the energy \cite{Jain:2007}, and to particles
which are confined to move on a spherical surface rather than on a
circle \cite{Nobre:2003}, have also shown the existence of QSS.
All these models evolve under deterministic Hamiltonian dynamics.
The robustness of quasistationarity to stochastic dynamical processes
has also been analyzed, where it is found that QSS exist only as a
crossover phenomenon. Under such dynamics, these states have a
finite relaxation time
which is determined by the rate of the stochastic
process \cite{Baldovin:2006,Baldovin:20061,
Baldovin:2009,Gupta:2010,Gupta:20101,Baldovin:2010}.

In this work, we address the issue of ubiquity of QSS in
long-range systems by examining a different dynamical model for its
existence than the HMF model. To this end, we consider an
anisotropic Heisenberg model with mean-field interactions. The model
comprises globally coupled three-component Heisenberg spins
evolving under classical spin dynamics. The dynamics is thus very different
from particle dynamics of previously studied long-range systems that have
shown the existence of QSS.

Our model has an equilibrium phase diagram with a continuous
transition from a low-energy magnetic phase to a high-energy
non-magnetic phase. Analysis of the Vlasov equation for evolution of
the phase space distribution shows that, as in previously studied
models, our model exhibits relaxation over times that grow either
logarithmically or algebraically with the system size. At low
energies, a non-magnetic initial state is dynamically unstable. It relaxes to the stable magnetically-ordered state on a logarithmic
timescale. At higher energies, but within the magnetic phase, the
non-magnetic state becomes linearly stable, and its relaxation to
equilibrium occurs on an algebraically growing timescale so that QSS
are observed. These results, obtained by analyzing the marginal
stability of the Vlasov equation, are supported by extensive
numerical simulations. This analysis yields further evidence for the
occurrence of QSS in generic long-range systems.

\section{The model}
\l{themodel}
We start by defining the model of study. It 
consists of $N$ globally coupled classical Heisenberg spins
of unit length, ${\mathbf S}_i=(S_{ix},S_{iy},S_{iz})$,
$i=1,2,\ldots,N$. In terms of spherical polar angles $\th_i \in
[0,\pi]$ and $\phi_i \in [0,2\pi]$ for the orientation of the $i$-th
spin, one has $S_{ix}=\sin \th_i \cos \ph_i, S_{iy}=\sin \th_i \sin
\ph_i, S_{iz}=\cos \th_i$. The Hamiltonian of the model is given by
\be
H=-\fr{J}{2N}\sum_{i,j=1}^N {\bf S}_i \cdot {\bf
S}_j+D\sum_{i=1}^N S^2_{iz},
\l{H}
\ee
where the first term with $J
> 0$ describes a ferromagnetic mean-field like coupling, while the
last term gives the energy due to a local anisotropy. We take $D>0$
such that at equilibrium, the energy is lowered by having the
magnetization ${\bf m}=(1/N)\sum_{i=1}^N{\bf S}_i$ pointing in the $xy$
plane. The coupling constant $J$ is scaled by $1/N$ to make the
energy extensive, in accordance with the Kac
prescription \cite{Kac:1963}. However, the system continues to
remain non-additive in the sense that it cannot be trivially divided into
independent macroscopic parts, as can be achieved with short-range
systems. In this work, we take $J=1$ and the Boltzmann constant
$k_B=1$.

The time evolution of the model (\ref{H}) is governed
by the set of first-order differential equations
\be
\fr{d{\bf S}_i}{dt}=\{{\bf S}_i,H\}; ~~~~i=1,2,\ldots,N.
\ee
Here the Poisson bracket $\{A,B\}$ for two functions of the spins
is obtained by noting that suitable canonical variables for a
classical spin are $\phi$ and $S_z$, so that in our model,
$\{A,B\}\equiv \sum_{i=1}^N (\partial A/\partial \phi_i
\partial B/\partial S_{iz}-\partial A/\partial S_{iz}
\partial B/\partial \phi_i)$. It may be rewritten as \cite{Mermin:1967}
\be
\{A,B\}=\sum_{i=1}^N {\bf S}_i \cdot \fr{\partial A} {\partial
{\bf S}_i}\times \fr{\partial B}{\partial {\bf S}_i}.
\ee
Using the
above relation, we obtain the equations of motion of the model:

\bea
&&\dot{S}_{ix}=S_{iy}m_z-S_{iz}m_y-2DS_{iy}S_{iz}, \l{eqnmotionx} \\
&&\dot{S}_{iy}=S_{iz}m_x-S_{ix}m_z+2DS_{ix}S_{iz}, \l{eqnmotiony}\\
&&\dot{S}_{iz}=S_{ix}m_y-S_{iy}m_x. \l{eqnmotionz}
\eea
Note that the above equations of motion may also be obtained by considering
the corresponding quantum equations and taking the limit of infinite spins.

From Eq. (\ref{eqnmotionz}), one finds by summing over $i$, that
$m_z$ is a constant of motion. The motion also conserves the total
energy and the length of each spin. From Eqs. (\ref{eqnmotionx}),
(\ref{eqnmotiony}), and (\ref{eqnmotionz}), the time evolution of
the variables $\th_i$ and $\ph_i$ are obtained as
\bea
\dot{\th_i}&=&m_x \sin \ph_i-m_y \cos \ph_i,\l{eqnmotiontheta}  \\
\dot{\ph_i}&=&m_x\cot \th_i \cos \ph_i+m_y \cot \th_i \sin \ph_i
-m_z+2D\cos \th_i. \l{eqnmotionphi}
\eea

Note that the Hamiltonian of the previously studied models of
particles moving either on a unit circle \cite{Ruffo:1995} or on a
spherical surface \cite{Nobre:2003} may also be expressed in terms
of spin variables. However, the dynamics of these models is rather
different from that of the model considered here. Unlike the present
model, all Poisson brackets of the spin variables of these models
are set to zero, and the dynamics is generated by a kinetic energy
term which is explicitly included in the Hamiltonians and which is
absent in the present model. It is worthwhile to point out that even in the
large $D$ limit, when the $z$-component of the spins becomes vanishingly
small, the dynamics is different from the HMF model with particles moving 
on a unit circle.

\section{Equilibrium phase diagram}
\l{equilibrium}
We now discuss the equilibrium phase diagram of our model.
The canonical partition function $Z$ is given by
\be Z=\int
\prod_i\sin \th_id\th_id\ph_i\exp\Big[\frac{\beta Nm^2}{2}- \beta
D\sum_i \cos^2\th_i\Big],
\l{Z}
\ee
where $\beta$ is the inverse temperature, and
$m=(m_x^2+m_y^2+m_z^2)^{1/2}$. Since Eq. (\ref{H}) describes a
mean-field system, it is straightforward to write down expressions
for equilibrium quantities like the average magnetization or the
average energy. As mentioned above, for $D>0$, the system orders in
the $xy$ plane. Choosing the ordering direction as $x$ without loss
of generality, the average equilibrium magnetization along $x$, given by
$\langle m_x \rangle=\langle \sin \th \cos \ph \rangle$, is
\be
\langle m_x \rangle=\fr{\int \sin^2 \th \cos \ph d\th d\ph
e^{\beta \langle m_x \rangle \sin \th \cos \ph-\beta D \cos^2 \th}}
{\int \sin \th d\th d\ph e^{\beta \langle m_x \rangle \sin \th \cos
\ph-\beta D \cos^2 \th}}. \l{mx}
\ee
Close to the critical point, one may expand the above transcendental
equation to leading order in $\langle m_x \rangle$ to get
\be
\langle m_x \rangle\Big( \int \sin \th d\th d\ph e^{-\beta_c D \cos^2 \th}-
\beta_c \int \sin^3\th \cos^2\ph d\th d\ph e^{-\beta_c D \cos^2
\th}\Big)=0.
\ee
The transition
temperature $\beta_c$ is obtained by equating the bracketed
expression to zero, and may be seen to satisfy
\be
\fr{2}{\beta_c
}=1-\fr{1}{2\beta_cD}+\fr{e^{-\beta_cD}}{\sqrt{\pi\beta_cD}
\mathrm{Erf}[\sqrt{\beta_cD}]}. \l{betac}
\ee
Here, $\mathrm{Erf}[x]=\fr{2}{\sqrt{\pi}}\int_0^x e^{-t^2}dt$ is the
error function. The critical energy density, $\eps_c=D\langle \cos^2
\th \rangle$, is
\be
\eps_c =D\left(1-\fr{2}{\beta_c }\right),
\l{ec}
\ee
see Fig. (\ref{betacfig}). In model (\ref{H}), where the phase transition is continuous, the
canonical and microcanonical ensembles are expected to be
equivalent \cite{Barre-book:2002,Bouchet:2005}, and Eq. (\ref{ec}) therefore also
yields the microcanonical
energy at the transition.

\section{Relaxation towards equilibrium}
\l{relaxation}
To study the dynamical behavior of magnetization, we now analyze
the Vlasov equation for evolution of the phase space
density \cite{Nicholson:1992}. For any mean-field model like ours, this
equation faithfully describes the
$N$-particle dynamics for
finite time and in the limit $N \rightarrow
\infty$ \cite{Braun:1977,review3}.
For the model (\ref{H}), we study the dynamics by examining a single particle,
which is moving in the two-dimensional phase space of its canonical coordinates, 
$\ph$ and $S_z=\cos \th$, due to the mean-field produced by all other particles. Here, 
and in the following, the
particle index is suppressed for brevity. Let $g(\cos \th, \ph,t)$ be the probability 
density in this single-particle phase space, such that $g(\cos \th,\ph,t)d\cos \th d\ph$ 
gives the probability to find the particle with the $z$-component of its spin between $\cos \th$ 
and $\cos \th+d\cos \th$ and the azimuthal angle between $\ph$ and $\ph+d\ph$ at time $t$.
In terms of the canonical coordinates, $\phi$ and $\cos \th$, the flow in the phase space is divergence free. 
Conservation of probability then implies vanishing of the total time derivative of $g$, that is,
\be 
\frac{dg}{dt}=\frac{\partial g}{\partial t}+\dot{\cos \th}\frac{\partial g}{\partial \cos \th}+\dot{\ph}\frac{\partial g}{\partial \ph}=0,
\ee 
where dot represents derivative with respect to time. More conveniently, we define a 
new function, $f(\th,\ph,t) \equiv g(\cos \th,\ph,t)$, 
such that $f(\th,\ph,t)\sin \th d\th d\ph$ gives the probability to find the particle 
with its angles between $\th$ and $\th+d\th$ 
and between $\ph$ and $\ph+ d\ph$ at time $t$. 
The equation for the evolution of $f$ is straightforwardly obtained from that of 
$g$ to yield $\partial f/\partial t+\dot{\th}(\partial f/\partial \th)+\dot{\ph}(\partial f/\partial \ph)=0$. 
Using Eqs. (\ref{eqnmotiontheta}) and (\ref{eqnmotionphi}) to substitute
for $\dot{\th}$ and $\dot{\ph}$ gives the Vlasov equation for time
evolution of $f(\th,\ph,t)$ as
\bea
\fr{\partial f}{\partial t}&=&\Big[m_y\cos \ph-m_x\sin \ph\Big]\fr{\partial f}{\partial \th}\nonumber \\
&-&\Big[m_x\cot \th\cos \ph+m_y\cot \th\sin \ph
-m_z+2D\cos \th\Big]\fr{\partial f}{\partial\phi}. 
\l{Vlasov}
\eea
In the above equation, the magnetization components are given by $(m_x,m_y,m_z)=\int
(\sin \th'\cos \ph',\sin \th' \sin \ph',\cos \th')f(\th',\ph',t)\sin\th'd\th'd\ph'$.

Now, consider an initial state prepared by sampling independently
for each of the $N$ spins the angle $\phi$ uniformly over $[0,2\pi]$
and the angle $\th$ uniformly over an arbitrary interval symmetric about
$\pi/2$. Such a
state will have the distribution
\be
f(\th,\ph,0)=\fr{1}{2\pi}p(\th),
\l{waterbag}
\ee
with $p(\th)$, the distribution for $\th$, given by
\be
p(\th)=\left\{
\begin{array}{ll}
               \fr{1}{2\sin a} & \mbox{if $\th \in \left[\fr{\pi}{2}-a,\fr{\pi}{2}+a\right]$}, \\
               & \\
               0 & \mbox{otherwise}.
               \end{array}
        \right. \\
\l{pth}
\ee
We call such an initial state a waterbag state, in analogy with a
similar form of an initial state studied in the context of the HMF model.
It is easily verified that this non-magnetic state has the energy
\be
\eps=\fr{D}{3}\sin^2 a,
\l{eps}
\ee
and that the state is stationary under the Vlasov dynamics (\ref{Vlasov}).
Let us proceed to analyze the dynamical stability of such a state. Such
stability analysis in the context of the HMF model was pursued
in \cite{Yamaguchi:2004}. Here, we closely follow the treatment
adopted in \cite{Jain:2007}. We consider finite
but large system size $N$, and linearize the Vlasov
equation (\ref{Vlasov}) with respect to finite-size
fluctuations $\delta f(\th,\ph,t)$ by expanding $f(\th,\ph,t)$ as
\be
f(\th,\ph,t)=\fr{1}{2\pi}p(\th)+\lambda \delta f(\th,\ph,t).
\l{perturbation}
\ee
Since the initial angles of the $N$ spins are sampled independently,
the small parameter
$\lambda$ is of order $1/\sqrt{N}$. After linearization,
Eq. (\ref{Vlasov}) yields the following equation for 
$\delta f(\th,\ph,t)$:
\be
\fr{\partial \delta f}{\partial t}=\Big[m_y\cos
\ph-m_x\sin \ph\Big]\fr{1}{2\pi}\fr{d p(\th)}{d
\th}-2D\cos \th\fr{\partial\delta f}{\partial\phi},
\l{linearVlasov}
\ee
where now $m_x$ and $m_y$ are linear in $\delta f$.

To study the linear dynamics, we note that at long times, it is
dominated by the mode corresponding to the largest eigenfrequency
$\o$ of the linearized equation (\ref{linearVlasov}). Since the
perturbation $\delta f(\th,\ph,t)$ is $2\pi$-periodic in $\ph$,
one has the following expansion in terms of the Fourier modes
$g_k(\th,\o)$, which is valid after a short initial transient 
(See \cite{Nicholson:1992}, Chapter 6):
\be
\delta f(\th,\ph,t)=\sum_k g_k(\th,\o)e^{i(k\ph+\o t)}.
\l{fexpansion}
\ee
The magnetization components are given by
\bea
m_x=\pi \int \sin^2 \th'd\th'\left(g_{-1}(\th',\o)+g_1(\th',\o)\right)e^{i\o
t}, \\
m_y=-i\pi \int \sin^2 \th'
d\th'\left(g_{-1}(\th',\o)-g_1(\th',\o)\right)e^{i\o t}. 
\eea
It thus follows that the relevant
eigenmodes of Eq. (\ref{linearVlasov}) have $k= \pm 1$. Using
Eq. (\ref{fexpansion}), and the above expressions for $m_x$ and $m_y$
in Eq. (\ref{linearVlasov}), we find that the coefficients
$g_{\pm 1}(\th,\o)$ satisfy
\be
g_{\pm 1}=\fr{d p(\th)}{d \th}
\fr{1}{2(2D\cos \th \pm \o)}\int\limits_{\pi/2-a}^{\pi/2+a}
g_{\pm 1}(\th', \o) \sin^2 \th' d\th'. \nonumber
\ee
On multiplying both sides of the above equation by $\sin^2 \th$
and then integrating over $\th$, we get
\be
I_\pm(1-K_\pm)=0,
\l{omegaeqn}
\ee
where
\be
I_{\pm}=\int_{\pi/2-a}^{\pi/2+a} g_{\pm 1}(\th, \o) \sin^2 \th d\th,
\ee
and
\be
K_\pm=\int_{\pi/2-a}^{\pi/2+a} \fr{d p(\th)}
{d \th} \fr{\sin^2 \th}{2(2D\cos \th \pm \o)}d\th.
\l{Kpm}
\ee
Since $I_\pm \ne 0$, it then follows from Eq. (\ref{omegaeqn}) that
the frequency $\o$ is given by the
condition
\be
K_\pm=1.
\l{Kpmcond}
\ee
From Eq. (\ref{pth}), we get
\be
\fr{d p(\th)}{d \th}
=\fr{1}{2\sin
a}\Big[\delta\Big(\th-\fr{\pi}{2}+a\Big)-\delta\Big(\fr{\pi}{2}+a-\th\Big)\Big],
\ee
which, together with Eqs. (\ref{Kpm}) and (\ref{Kpmcond}), give the following expression for
the largest eigenfrequency $\o$:
\be
\o^2=4D^2 \sin^2 a - D\cos^2 a.
\ee
Expressing the parameter $a$ in the above equation in terms of
the energy $\eps$ in Eq. (\ref{eps}) finally yields
\be
\o^2=\eps(3+12D)-D.
\l{omegasquared}
\ee
\bef
\begin{center}
\includegraphics[width=100mm]{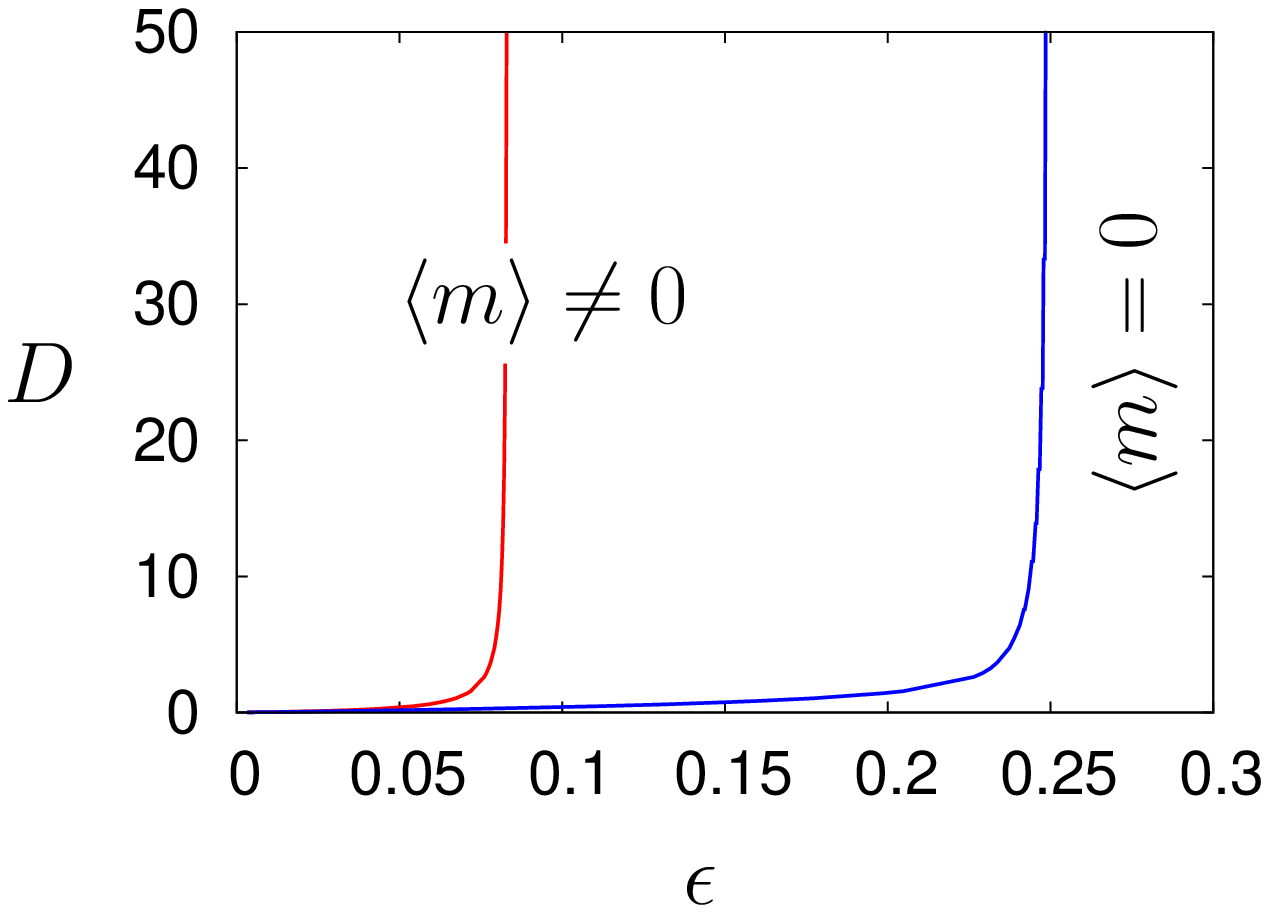}
\caption{Dynamical phase diagram of our model in the $(\eps,D)$
plane for the choice of the waterbag initial state, Eq. (\ref{waterbag}).
The line to the right is the thermodynamic phase boundary
$D(\eps_c)$ between the magnetic and the non-magnetic phase, and is
given by Eq. (\ref{ec}) with $\beta_c$ obtained by solving Eq. (\ref{betac}).
The line to the left gives the dynamical phase boundary $D(\eps^*)$,
and is given by Eq. (\ref{estar}).}
\l{betacfig}
\end{center}
\eef

We thus see that the frequency $\o$ is real for $\eps > \eps^*$, given by
\be
\eps^*=\fr{D}{3+12 D},
\l{estar}
\ee
see Fig. (\ref{betacfig}). Therefore, unstable modes do
not exist in this energy range so that the waterbag state (\ref{waterbag}) is linearly stable. Hence, 
QSS is observed. In this
case, in a finite system, such a state eventually relaxes to
Boltzmann-Gibbs equilibrium on a timescale over which non-linear
correction terms should be added to the Vlasov equation \cite{review3}.

On the other hand, for $\eps < \eps^*$, the waterbag state is unstable.
Consequently, the perturbation $\delta f(\th,\ph,t)$
grows exponentially fast towards Boltzmann-Gibbs equilibrium. Below $\eps^*$,
on setting $\o^2=-\O^2$ with real $\O$,
we get $\delta f(\th,\ph,t) =Ae^{\pm i\ph + \O t}$, where
$A$ is a constant. Consequently, the average
magnetization behaves as
\be
m(t) \sim \fr{1}{\sqrt{N}}e^{\O t}; ~~~~~~~~~~~~\eps < \eps^*,
\l{epsltepsstar}
\ee
before it relaxes to the equilibrium value.
From the above equation, it follows that for $\eps < \eps^*$,
the relaxation timescale $\tau(N)$
over which the magnetization acquires the equilibrium value of
$O(1)$ scales as $\ln N$.

\section{Numerical simulations}
\l{numerics}
\bef
\begin{center}
\includegraphics[width=100mm]{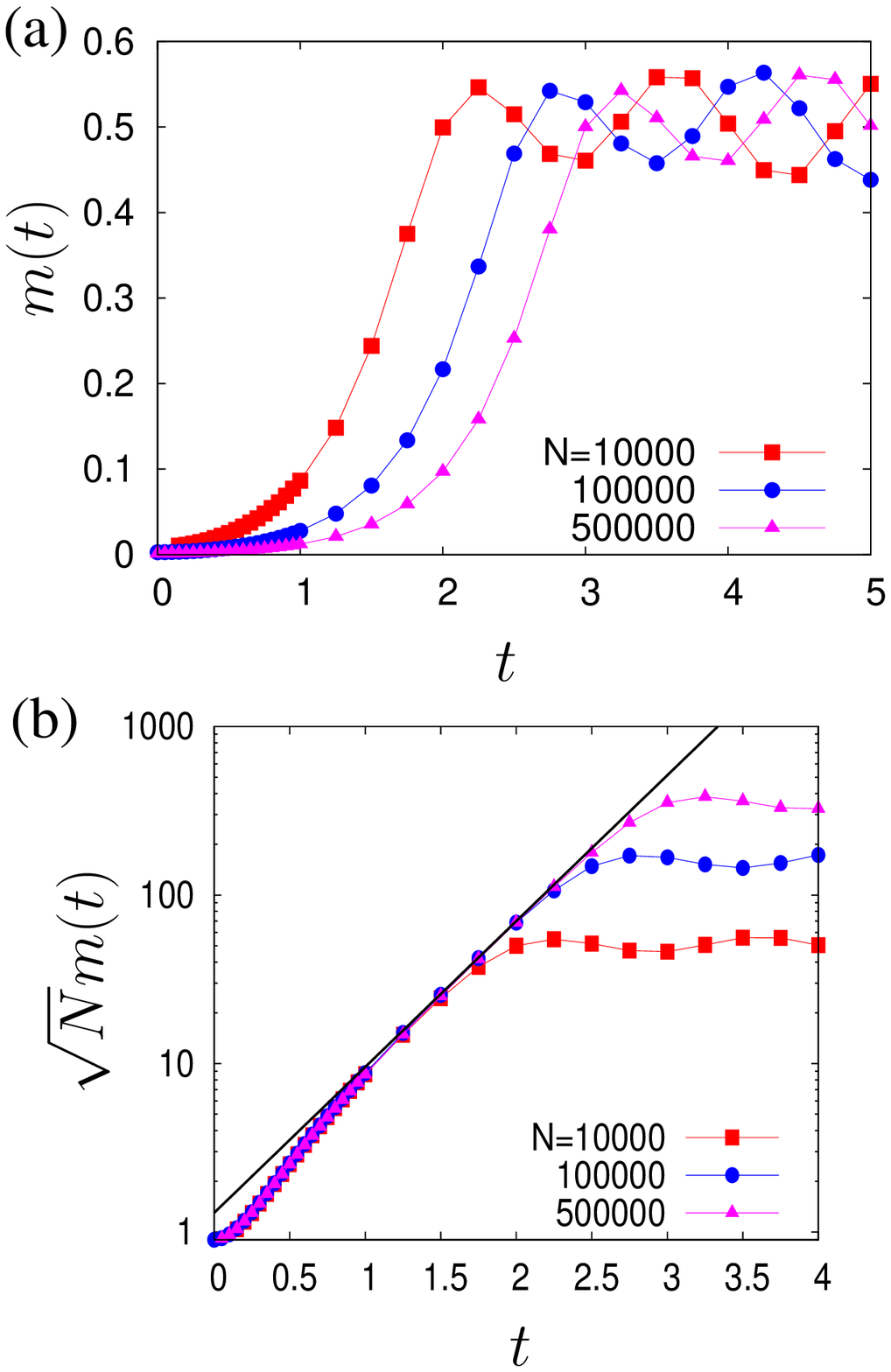}
\caption{(a) Time evolution of average magnetization $m(t)$
in the unstable phase with energy density $\eps=0.0603$, the parameter $D=15$,
and for systems of size $N=10000,100000,500000$ (left to right).
(b) Data collapse for the
scaled magnetization $\sqrt{N}m(t)$ as a function of $t$, in
accordance with Eq. (\ref{epsltepsstar}). The black line represents
the function $s(t) \sim e^{\O t}$, with $\O$ obtained from
Eq. (\ref{omegasquared}) by
substituting $\o^2=-\O^2$. Data averaging varies between $2 \times 10^4$
histories for the smallest system and $500$ histories for the largest one.}
\l{elt}
\end{center}
\eef
\bef
\begin{center}
\includegraphics[width=100mm]{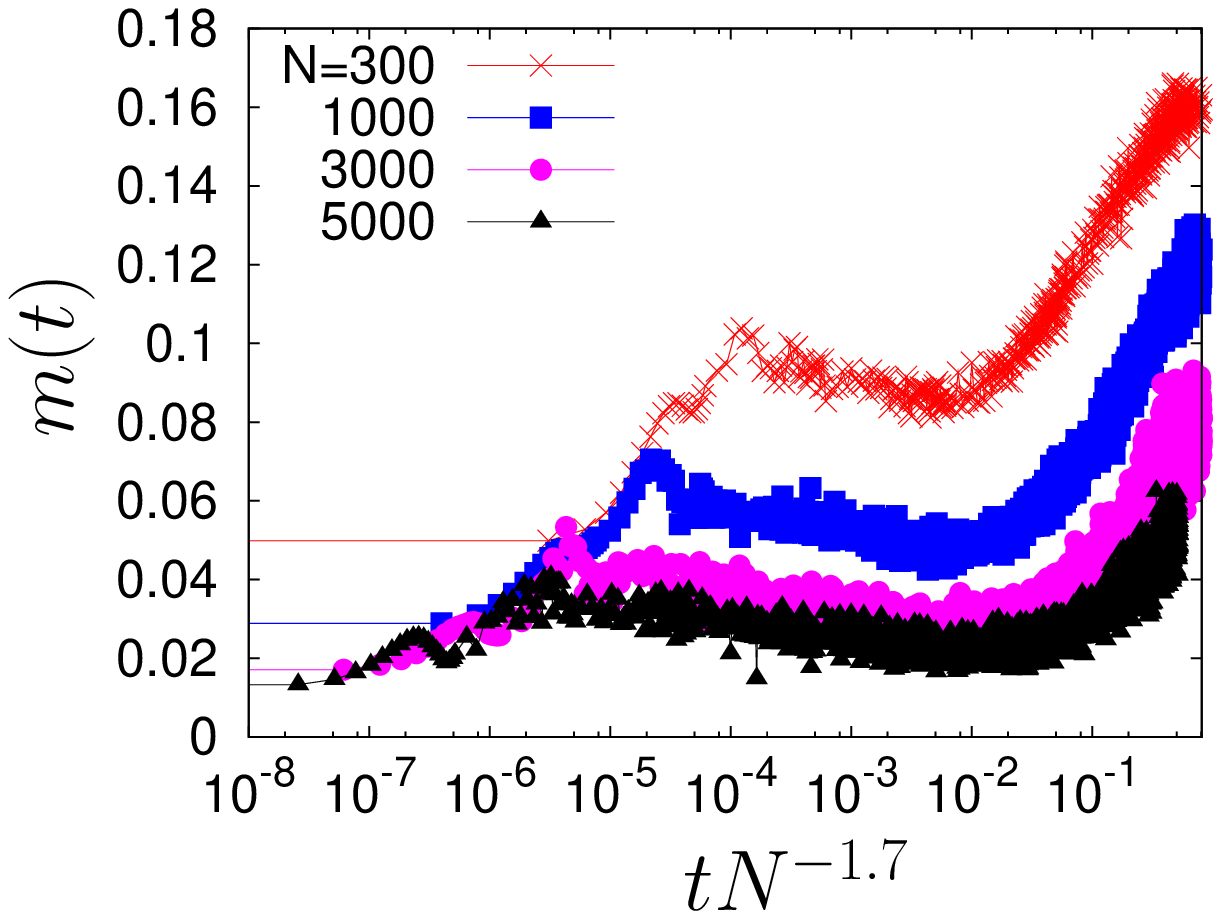}
\caption{Average magnetization $m(t)$ as
a function of $tN^{-1.7}$ in the stable phase with energy density
$\eps=0.24$, the parameter $D=15$, and for systems of size $N=300,1000,3000, 5000$ (top to
bottom).
Data averaging varies between $300$ histories for the smallest
system and $25$ histories for the largest one. The figure suggests a QSS
life-time $\tau(N) \sim N^{1.7}$.}
\l{egt}
\end{center}
\eef
In order to verify these features, we performed numerical
simulations of the dynamics by integrating the equations of
motion (\ref{eqnmotionx}), (\ref{eqnmotiony}), (\ref{eqnmotionz})
by using a fourth-order Runge Kutta method with time step equal
to $0.01$. For $\eps < \eps^*$, the results presented in Fig. \ref{elt}(a)
show that the magnetization grows fast towards equilibrium. On scaling the
magnetization by $\sqrt{N}$,
Fig. \ref{elt}(b) shows a very good scaling collapse in accordance with the
exponential growth predicted by Eq. (\ref{epsltepsstar}). The growth
rate $\O$ is in agreement with that obtained from Eq. (\ref{omegasquared}) by
substituting $\o^2=-\O^2$.
For energies $\eps^* < \eps < \eps_c$, when the waterbag state (\ref{waterbag})
is linearly stable, Fig. \ref{egt} suggests a much longer
relaxation time $\tau(N) \sim N^{1.7}$. A similar scaling of
the QSS relaxation time was also observed in the HMF
model \cite{Yamaguchi:2004}.

\section{Conclusions}
\l{conclusions}In conclusion, we addressed the ubiquity of non-Boltzmann quasistationary
states (QSS) during relaxation of long-range systems. This was done by studying an
anisotropic Heisenberg model of globally coupled classical 
spins evolving under classical spin dynamics. Quasistationary states have earlier been shown
to occur in long-range interacting systems composed of particles
(inertial rotors) which
are evolving under particle dynamics dictated by the underlying Hamiltonian.
Thus, our model provides a different possible setting for the occurrence of
QSS under spin dynamics. By analyzing the Vlasov equation for the time
evolution of the phase space distribution, we demonstrated that in this model,
relaxation of a class of initial states in certain energy interval proceeds
through intermediate QSS. These states have a lifetime that grows algebraically with the
system size. This further establishes the possibility of long-range systems to
exhibit quasistationarity under a broader class of dynamical processes.

\section{Acknowledgments}
We thank A. Bar, O. Cohen, T. Dauxois, O. Hirschberg, 
S. Levit and S. Ruffo for helpful discussions and comments 
on the manuscript. The
support of the Israel Science Foundation (ISF) and the Minerva
Foundation with funding from the Federal German Ministry for
Education and Research is gratefully acknowledged.

\section*{References}

\end{document}